\documentclass[aps,preprint]{revtex4}%
\usepackage{amsfonts}
\usepackage{amsmath}
\usepackage{amssymb}
\usepackage{graphicx}%
\providecommand{\U}[1]{\protect\rule{.1in}{.1in}}

\begin{document}
\author{}
\title{Emergence of cosmic space and horizon thermodynamics in the context of the quantum-deformed entropy}
\author{Jianming Chen}
\email{jianming379@163.com}
\author{Gerui Chen}
\thanks{Corresponding author}
\email{chengerui@163.com}
\affiliation{College of Electronic Information and Physics,
\\ Central South University of Forestry and Technology, Changsha 410004, China}

\begin{abstract}
According to the quantum deformation approach to quantum gravity, the thermodynamical entropy of a quantum-deformed (q-deformed) black hole with horizon area $A$ established by Jalalzadeh is expressed as $S_q = \pi\sin \left( \frac{A}{8G\mathcal N} \right) /\sin\left(\frac{\pi}{2\mathcal N}\right)$, where $\mathcal N=L_q^2/L_{p}^2$ is the q-deformation parameter, $L_{p}$ denotes the Planck length, and $L_q$ denotes the quantum-deformed cosmic apparent horizon distance.
In this paper, assuming that the q-deformed entropy is associated with the apparent horizon of the Friedmann-Robertson-Walker (FRW) universe, we derive the Friedmann equation from the unified first law of thermodynamics, ${dE = T_q dS_q + WdV}$. And this one obtained is in line with the Friedmann equation derived from the law of emergence proposed by Padmanabhan. It clearly shows the connection between the law of emergence and the unified first law of thermodynamics.
Subsequently, in the context of the q-deformed horizon entropy, we investigate the constraints of entropy maximization, and result demonstrates the consistency of the law of emergence with the maximization of the q-deformed horizon entropy. Hence, the law of emergence can be understood as a tendency to maximize the q-deformed horizon entropy.
\end{abstract}

 \maketitle

\section{Introduction}\label{Intro}
The discovery of Newton's law of universal gravitation, one of the greatest achievements of natural science in the 17th century, reveals the laws of celestial motion. As the most universal force of nature, the origin of gravity is still being explored.
Einstein believes that gravity is just the curvature of space-time.
In the long history of the exploration of gravity, people have been trying to reveal its essence.

Jacobson \cite{Jacobson}, who investigated the thermodynamics of space-time and demonstrated clearly that Einstein's equation of general relativity is merely an equation of state for space-time, made a significant contribution to this field.
Since Jacobson did the research, many investigations have been carried out to reveal the intricate relationship between thermodynamics and gravity \cite{Eling,Cai1,Cai11,Padmb1,Padmb11,Padmb12,Padmb2,Padmb21,Padmb22}. The researches were then gradually extended to cosmic models \cite{Cai2,Cai Kim,Cai3,Frolov,PVer,PVer1,Cai4,Sheykhi1,Sheykhi2}, showing that the Friedmann equations of the Friedmann-Robertson-Walker (FRW) universe can be expressed as a form of conservation of energy on the apparent horizon.
Jacobson also revealed that the deep connection between the law of gravity and the laws of thermodynamics is motivated by the thermodynamics of the black hole. The correspondence between the two is often referred to as the thermodynamic-gravitational conjecture \cite{K1}.

In 2011, Verlinde \cite{Verlinde1} proposed that gravity is not a fundamental force but an entropic force, which is understood as the entropy-changing force of gravity caused by information on the holographic screen. From the equipartition theorem and the holographic principle, Verlinde deduced Newton's law of universal gravitation, Poisson's equation, and the Einstein field equations of the relativistic system \cite{Verlinde1}. Nowadays, studies of the origin of gravitational entropy have been extended to a certain extent in different contexts \cite{Cai5,Cai6,Origin1,Sheykhi3,Origin2,YCai,MLi,Sheykhi4,Sheykhi41}. Verlinde's proposal, viewing the gravitational field equations as the result of the changes in entropy caused by information on a holographic screen, enriches our understanding of the origin of gravity.
It is notable that the gravitational field equation is derived from thermodynamic arguments, so any modification in the entropy expression will modify the final corresponding field equation. The expression of entropy is the decisive factor, and plays a crucial role \cite{Sheykhi5,K3,K4,K41,K5,K51}.

Soon after, in 2012, Padmanabhan \cite{Padmb3} came up with a new viewpoint for the emergence of spatiotemporal dynamics to address questions by considering the emergence of space-time in cosmology. Padmanabhan found that the spatial expansion of the universe could be considered a consequence of the emergence of space as cosmic time progresses. Padmanabhan derived the Friedmann equation of the flat FRW universe by calculating the difference in degrees of freedom between the boundary and the bulk of the universe \cite{Padmb3}.
Considering a spatially flat universe, Cai \cite{Cai7} extended the procedure to higher-dimensional theories of gravities (Gauss-Bonnet and more general Lovelock gravities) by making appropriate modifications to surface degrees of freedom \cite{Cai7}. Padmanabhan's ideas had attracted widespread attention \cite{Other1,Other2,Sheykhi,Other3,Other4,Other5}. By studying and refining Padmanabhan's proposal, one can extract the Friedmann equation of the FRW universe with any spatial curvature, and there are many relevant studies currently \cite{Chen,Study1,Study2}.

As is known to all, every ordinary macroscopic system eventually evolves into an equilibrium state with maximum entropy \cite{Max1}. Pavon and Radicella have shown that our universe has a history of Hubble expansion, which manifests itself as an ordinary macroscopic system \cite{Max2}.
Later on, Krishna and Mathew explored the relationship between the emergence of cosmic space and the entropy maximization of the horizon \cite{Max3,Max4,Max5}.
In the context of general relativity, the holographic equipartition law proposed in Ref.\cite{Padmb3} effectively verified the maximization of entropy \cite{Max3}.
In Ref.\cite{Cai7} and Ref.\cite{Max6}, it has been shown that the generalized holographic equipartition under Einstein, Gauss-Bonnet and Lovelock gravities is consistent with the constraints of entropy maximization \cite{Max4}. It should be noted that the results in Ref.\cite{Max3} and Ref.\cite{Max4} are limited to the flat universe.
However, in a non-flat universe, the consistency of the constraints of entropy maximization and the generalized holographic equipartition has also been verified in Ref.\cite{Max5}.
Taken together, Krishna and Mathew found that the law of emergence and the horizon entropy maximization result in the same constraints, so the emergence of cosmic space can be seen as a trend towards horizon entropy maximization.

There are currently a number of well-known definitions of entropy, such as the Tsallis entropy \cite{Tsallis1,Tsallis2}, the Renyi entropy \cite{Renyi}, the Barrow entropy \cite{Barrow1}, the Kaniadakis entropy \cite{Kaniad1,Kaniad2}, and the entropy in the setting of loop quantum gravity \cite{Loop1}. The investigation of the entropy for black hole in cosmology has never ceased.
At present, researches in the field of quantum are very popular, and quantum intelligence and computations are applied in many research fields.
According to various quantum gravity proposals, Jalalzadeh \cite{Quan1} has recently proposed a new entropy formula for the black hole based on the quantum deformation approach to quantum gravity. This entropy, referred to as the q-deformed entropy (quantum-deformed entropy), is defined as \cite{Quan0}
\begin{eqnarray}\label{q0 entropy}
S_q=\pi \frac {\sin(\frac{A}{8G\mathcal N})}{\sin(\frac{\pi}{2\mathcal N})},
\end{eqnarray}
where $A$ is the black hole horizon area, and ${\mathcal N}$ is the q-deformation parameter given by ${\mathcal N}={L^2_{q}}/{{L}^2_p} $. Similar to the Planck length ${L}_p$, $L_{q}$ is a fundamental dimensional constant of quantum gravity theory. The q-deformation is thought to be related to a fundamental dimensional constant.
Simply speaking, in terms of the ladder operators $\{a_-, a_+\}$ and the deformation parameter ${q}$, new expressions for the area of the event horizon and the mass of the black hole are obtained, which can lead us to derive the entropy. The deformation parameter ${q}$, is a primitive root of unity \cite{Quan0}.
The q-deformed entropy takes the form of a fraction, and both the numerator and denominator are trigonometric functions.
Furthermore, at the classical limit of quantum geometry, $\mathcal N \rightarrow \infty$, the area law is restored.

This research aims to study the effects of the law of emergence in the context of the q-deformed entropy. The Friedmann and Raychaudhuri equations based on q-deformed entropy have been partially explored in Ref.\cite{Quan0}. Jalalzadeh found that the equations indicate an effective dark energy component, which is able to explain the late-time acceleration of the universe.
Our work differs from Ref.\cite{Quan0} in that we employ the unified first law of thermodynamics to arrive at the modified Friedmann equation as the universe is expanding. Then, we investigate the consistency of the constraints of the generalized holographic equipartition and horizon entropy maximization. In the final step, we verify whether the law of emergence leads to the maximization of the q-deformed horizon entropy.

The remainder of this paper is organized as follows.
In Sect.\ref{FirstLaw}, we derive the modified Friedmann equation from q-deformed entropy at the apparent horizon of a $(3+1)$-dimensional FRW universe based on the unified first law of thermodynamics, ${dE=T_q dS_q+W dV}$.
In Sect.\ref{Emerg}, employing the modified version of the law of emergence of cosmic space, we obtain the Friedmann equation from q-deformed entropy, which is the same as the one in the previous section.
In Sect.\ref{Max}, after obtaining the constraints of the maximization of q-deformed horizon entropy, we examine the consistency between the law of emergence and the horizon entropy maximization in the context of q-deformed entropy.
The last section is devoted to conclusions and discussions.
In this paper, $k_{B}=c=\hbar=1 $ has been set for simplicity.

\section{Modified Friedmann equation from the unified first law of thermodynamics at the apparent horizon}\label{FirstLaw}
Assumed to be spatially homogeneous and isotropic, the background of space-time is represented by the line element as
\begin{eqnarray}
ds^2={h}_{\mu \nu}dx^{\mu}
dx^{\nu}+\tilde{r}^2(d\theta^2+\sin^2\theta d\phi^2),
\end{eqnarray}
where $\tilde{r}=a(t)r$, $x^0=t, x^1=r$, the 2-dimensional metric ${h_{\mu\nu} = \text{diag}(-1, a^2/(1-kr^2))}$, and $a(t)$ is the time-dependent scale factor.
Here $k$ denotes the curvature of space, and $k = -1, 0, 1$ correspond respectively to the open, flat, and closed universes. The physical boundary of the universe, which is consistent with laws of thermodynamics, is the apparent horizon with radius
\begin{eqnarray}\label{radius}
\tilde{r}_A=\frac{1}{\sqrt{H^2+k/a^2}},
\end{eqnarray}
where $H = \dot{a}/a$ is the Hubble parameter, and $a(t)$ is abbreviated to $a$.

In the very beginning, the radius of the black hole is replaced with the apparent horizon radius ${\tilde{r}_A}$, so $A$ is the apparent horizon area of the universe.
Thus, the q-deformed entropy of the apparent horizon is rewritten as
\begin{eqnarray}\label{q entropy}
S_q=\pi\frac{\sin(\frac{\lambda}{G} \tilde{r}_A^2)}{\sin(\lambda)}, ~0\leq \frac{\lambda}{G} \tilde{r}_A^2\leq\frac{\pi}{2},
\end{eqnarray}
where $\lambda=\pi/(2\mathcal N)
=\frac{\pi}{2}({L_{p}}/{L_q})^2
=\frac{\pi}{6}G\Lambda_q$.
It considers that $L_q \equiv \sqrt{3/\Lambda_q}$, where $\Lambda_q$ is the q-cosmological constant.

We consider the universe's matter and energy content to be a perfect fluid with a stress-energy tensor
\begin{eqnarray}\label{T1}
T_{\mu\nu}=(\rho+p)u_{\mu}u_{\nu}+pg_{\mu\nu},
\end{eqnarray}
where $\rho$, $p$, and $u_\mu$ are the energy density, pressure, and 4-velocity field of the fluid, respectively.
The conservation of the energy-stress tensor in the FRW background, $\nabla_{\mu}T^{\mu\nu}=0$, leads to the continuity equation
\begin{eqnarray}\label{Cont}
\dot{\rho}+3H(\rho+p)=0.
\end{eqnarray}

The definition of the temperature linked to the apparent horizon in Ref.\cite{Cai2} is given by
\begin{eqnarray}\label{T}
T_q=\frac{\kappa}{2\pi}=-\frac{1}{2 \pi \tilde
r_A}\left(1-\frac{\dot {\tilde r}_A}{2H\tilde r_A}\right),
\end{eqnarray}
where $\kappa$ is the $\text{Kodama}$ surface gravity, and the dot over $\tilde r_A$ denotes the derivative with respect to the cosmic time $t$.
It shows that $T\leq0$ at the time when $\dot {\tilde r}_A \leq 2H\tilde r_A$. To make sure that the temperature is not negative, it should have $T=|\kappa|/(2\pi)$. We assume $\dot {\tilde r}_A\ll 2H\tilde r_A$ within an infinitesimal internal of time $dt$, which physically means that the apparent horizon radius ${\tilde{r}_A}$ is kept fixed \cite{Cai Kim}.
Thus, the temperature is simplified to
\begin{eqnarray}\label{T0}
T_q = \frac {1}{2\pi\tilde r_A }.
\end{eqnarray}
Following Ref.\cite{SSM}, the work is defined as
\begin{eqnarray}\label{Work}
W=-\frac{1}{2} T^{\mu\nu}h_{\mu\nu}.
\end{eqnarray}
With simple calculation, it becomes
\begin{eqnarray}\label{Work2}
W=\frac{1}{2}(\rho-p).
\end{eqnarray}
The work density term is caused by the change in the apparent horizon radius, which is actually about the volume change of the universe. The unified first law of thermodynamics is assumed to be satisfied at the apparent horizon, which is given by \cite{Cai2}
\begin{eqnarray}\label{FL}
dE = T_q dS_q + W dV.
\end{eqnarray}
From observing the form of the equation, we see that it clearly resembles the standard first law of thermodynamics. For a pure de Sitter space where $\rho=-p$, the work (\ref{Work2}) becomes $-p$, and the unified first law of thermodynamics (\ref{FL}) reduces to the standard form, ${dE = TdS - pdV}$.

The energy $E=\rho V$ is supposed to be the total energy content of the universe inside a sphere of radius $\tilde{r}_{A}$, where $V = 4\pi\tilde{r}_{A}^{3}/{3}$ is the volume enveloped by a $3$-dimensional sphere with the area $A=4\pi\tilde {r}_{A}^{2}$.
Hence, the total energy content of the universe
\begin{eqnarray}\label{E1}
E=\rho \frac{4\pi}{3} \tilde {r}_{A}^{3}.
\end{eqnarray}
Taking the differential form of $E$, we obtain
\begin{eqnarray}\label{dE1}
dE=4\pi\tilde {r}_{A}^{2} \rho d\tilde {r}_{A} + \frac{4\pi}{3}\tilde{r}_{A}^{3}\dot{\rho} dt.
\end{eqnarray}
By the continuity equation (\ref{Cont}), the above equation becomes
\begin{eqnarray}\label{dE2}
dE=4\pi\tilde {r}_{A}^{2}\rho d\tilde {r}_{A} -
4\pi H\tilde{r}_{A}^{3}(\rho+p) dt.
\end{eqnarray}
It could be assumed that the entropy related to the apparent horizon is in the form of q-deformed entropy.
Differentiating the q-deformed entropy modified by apparent horizon radius (\ref{q entropy}), we obtain
\begin{eqnarray} \label{dS}
dS_q = \frac{2\pi \lambda}{G}
\frac{\cos(\frac{\lambda}{G} \tilde {r}_{A}^2)} {\sin(\lambda)}
\tilde {r}_{A} d\tilde {r}_{A}.
\end{eqnarray}
Substituting Eq.(\ref{T0}), Eq.(\ref{Work2}), Eq.(\ref{dE2}) and Eq.(\ref{dS}) in the unified first law of thermodynamics (\ref{FL}), we get
\begin{eqnarray} \label{Fried1}
\frac{\lambda}{4\pi G}
\frac{\cos(\frac{\lambda}{G} \tilde {r}_{A}^2)} {\sin(\lambda)} \frac{1}{\tilde{r}_{A}^3}d\tilde {r}_{A}
= H(\rho+p) dt.
\end{eqnarray}
Applying the continuity equation (\ref{Cont}), we reach
\begin{eqnarray} \label{Fried2}
- \frac{\lambda}{4\pi G} \frac{\cos(\frac{\lambda}{G} \tilde {r}_{A}^2)} {\sin(\lambda)} \frac{1}{\tilde {r}_{A}^3} d\tilde {r}_{A}
= \frac{1}{3} d\rho.
\end{eqnarray}
By integrating Eq.(\ref{Fried2}), we get
\begin{eqnarray} \label{Fried3}
- \int{\frac{\cos(\frac{\lambda}{G} \tilde {r}_{A}^2)}{\tilde {r}_{A}^3} d\tilde {r}_{A}}
= \frac{\sin(\lambda)}{\lambda} \int{ \frac{4\pi G}{3} d\rho},
\end{eqnarray}
which can be written as
\begin{eqnarray} \label{Fried4}
\frac{\cos(\frac{\lambda}{G} \tilde {r}_{A}^2)}{\tilde {r}_{A}^2} +
\frac{\lambda}{G} Si(\frac{\lambda}{G} \tilde {r}_{A}^2)
= \frac{\sin(\lambda)}{\lambda} \frac{8\pi G}{3} \rho,
\end{eqnarray}
where the integration constant has been set to zero, and
\begin{eqnarray} \label{Si(x)}
Si(x)=\int_0^x\frac{\sin(t)}{t}dt
\end{eqnarray}
is the integral sine function.

With the value of $ \lambda = \frac{\pi}{2} ({L_{p}}/{L_q})^2 \simeq 10^{-123} $, Eq.(\ref{Fried4}) can be simplified to
\begin{eqnarray}\label{Fried final}
\frac{\cos(\frac{\lambda}{G}\tilde {r}_{A}^2)}{\tilde {r}_{A}^2} +\frac{\lambda}{G} Si(\frac{\lambda}{G}\tilde {r}_{A}^2)
= \frac{8\pi G}{3}\rho,
\end{eqnarray}
which is the modified Friedmann equation based on the q-deformed entropy.

Note that at the classical limit of quantum geometry, $\mathcal N \rightarrow \infty$ (that is, ${\lambda}\rightarrow0$), using relation (\ref{radius}), we discover that Eq.(\ref{Fried final}) reduces to the standard Friedmann equation
\begin{eqnarray}\label{Fried origin}
H^2+\frac{k}{a^2} = \frac{8\pi L_p^2}{3}\rho ,
\end{eqnarray}
where $\rho=\rho_\text{c}+\rho_\text{rad}$ is the fluid’s total energy density, $\rho_\text{c}$ is the energy density of cold matter, and $\rho_\text{rad}$ is the energy density of the radiation. Besides, $G$ is replaced with $L_p^2$, based on the Planck length $L_{p}=1/M_\text{P}$ and the Planck mass $M_\text{P}=1/\sqrt{G}$.

Supposing that the apparent horizon area includes quantum deformed characteristics, we apply the unified first law of thermodynamics, ${dE = T_q dS_q + WdV}$, at the apparent horizon of the FRW universe to get the corresponding modified Friedmann equation of the FRW universe.

\section{Modified Friedmann equation from the emergence of cosmic space}\label{Emerg}
Padmanabhan regards the spatial expansion of our universe as the consequence of the emergence of space, and believes that cosmic space emerges with progress in cosmic time \cite{Padmb3}. Moreover, the emergence of cosmic space is related to the difference between the number of degrees of freedom on the holographic surface and the one in the bulk. According to this proposal, Padmanabhan argued that in an infinitesimal interval $dt$ of cosmic time, the increase $dV$ of the cosmic volume is given by
\begin{eqnarray}\label{dV}
\frac{dV}{dt}=L_{p}^{2}(N_{\rm{sur}}-N_{\rm{bulk}}),
\end{eqnarray}
where $N_{\rm{sur}}$ is the number of degrees of freedom on the holographic surface, and $N_{\rm{bulk}}$ is the number of degrees of freedom in the bulk. However, this proposal is based on a flat universe. After Padmanabhan successfully derived the Friedmann equation of a flat FRW universe, Cai explored the Friedmann equations of a non-flat FRW universe in some other gravity theories, such as Gauss-Bonnet and Lovelock gravities \cite{Cai7}.
Sheykhi argued that in a non-flat universe, the proposal in Eq.(\ref{dV}) should be generalized to \cite{Sheykhi6}
\begin{eqnarray}\label{dV1}
\frac{dV}{dt}=L_{p}^{2}\frac{\tilde{r}_A}{H^{-1}}
\left(N_{\rm{sur}}-N_{\rm{bulk}}\right).
\end{eqnarray}
It implies that, in a non-flat universe, similar to a flat universe, the increase in volume is still proportional to the difference between the number of degrees of freedom on the boundary and in the bulk.
The distinction is that the function of proportionality is equal to the ratio of the apparent horizon and Hubble radius in a non-flat universe. For the spatially flat universe, $\tilde{r}_A =H^{-1}$, evidently it can recover the proposal (\ref{dV}).

Assuming the q-deformed entropy is associated with the apparent horizon, we are going to derive the modified Friedmann equation from the proposal of Eq.(\ref{dV1}).
First of all, the effective area of the holographic surface corresponding to the q-deformed entropy (\ref{q entropy}) is supposed to be
\begin{eqnarray}
\widetilde{A} = 4\pi L_p^2 \frac{\sin(\frac{\lambda}{G} \tilde{r}_A^2)}{\sin(\lambda)}.
\end{eqnarray}
Next, the increase in the effective volume satisfies
\begin{eqnarray}\label{dVt1}
\frac{d\widetilde{V}}{dt}& = &\frac{\tilde{r}_A}{2}\frac{d\widetilde{A}}{dt}
= 4\pi L_p^2 \frac{\lambda}{G} \frac{\cos(\frac{\lambda}{G} \tilde{r}_A^2)}{\sin(\lambda)} \tilde{r}_A^2 \dot{\tilde{r}}_A \\
&=& -2\pi \frac{\lambda}{\sin(\lambda)}\tilde{r}_A^5  \frac{d}{dt} \left[\frac{\cos(\frac{\lambda}{G} \tilde{r}_A^2)}{\tilde{r}_A^2} + \frac{\lambda}{G} Si(\frac{\lambda}{G}\tilde{r}_A^2) \right]. \notag
\end{eqnarray}
Inspired by the similar methods in Ref.\cite{Chen} and Ref.\cite{Study2}, we suppose that the number of degrees of freedom on the apparent horizon is given by
\begin{eqnarray}\label{Nsur2}
N_{\rm{sur}} &=& -2\pi \frac{\lambda}{\sin(\lambda)}\tilde{r}_A^4 \left[\frac{\cos(\frac{\lambda}{G} \tilde{r}_A^2)}{\tilde{r}_A^2} + \frac{\lambda}{G} Si(\frac{\lambda}{G}\tilde{r}_A^2)\right](-\frac{2}{L_p^2}) \nonumber\\
&=& \frac{4\pi}{L_p^2} \frac{\lambda}{\sin(\lambda)} \tilde{r}_A^4
\left[\frac{\cos(\frac{\lambda}{G} \tilde{r}_A^2)}{\tilde{r}_A^2} + \frac{\lambda}{G} Si(\frac{\lambda}{G}\tilde{r}_A^2)\right].
\end{eqnarray}
The temperature associated with the apparent horizon is the Hawking temperature \cite{Cai Kim}, which is given by
\begin{eqnarray}\label{T2}
T=\frac{1}{2\pi \tilde{r}_A}.
\end{eqnarray}
The ${\rm{Komar}}$ energy contained inside the sphere with volume $V=4 \pi \tilde{r}^3_A /3$ is
\begin{eqnarray}\label{Komar}
E_{\rm{Komar}}=|(\rho +3p)|V.
\end{eqnarray}
According to the equipartition law of energy, the number of degrees of freedom in the bulk is defined as
\begin{eqnarray}
N_{\rm{bulk}}=\frac{2E_{\rm{Komar}}}{T}.
\end{eqnarray}
Considering that $\rho+3p<0$, which is consistently with the condition of an accelerated universe \cite{Study2}, we can remove the absolute value sign in Eq.(\ref{Komar}).
Thus, the bulk degrees of freedom is obtained as
\begin{eqnarray}\label{Nbulk}
N_{\rm bulk}=-\frac{16 \pi^2}{3}  \tilde{r}_A^{4} (\rho+3p).
\end{eqnarray}
We replaced $V$ with $\widetilde{V}$ in proposal (\ref{dV1}), and rewrite it as
\begin{eqnarray}\label{dV2}
\frac{d\widetilde{V}}{dt}= L_p^2\frac{\tilde{r}_A}{H^{-1}}
(N_{\rm{sur}}-N_{\rm{bulk}}).
\end{eqnarray}
Substituting Eq.(\ref{dVt1}), Eq.(\ref{Nsur2}) and Eq.(\ref{Nbulk}) in relation (\ref{dV2}), after simplifying, we obtain
\begin{eqnarray}\label{Frgb1}
 \frac{\cos(\frac{\lambda}{G} \tilde{r}_A^2)}{\tilde{r}_A^3} \frac{\dot{\tilde{r}}_A}{H} - \left[\frac{\cos(\frac{\lambda}{G} \tilde{r}_A^2)}{\tilde{r}_A^2} + \frac{\lambda}{G} Si(\frac{\lambda}{G} \tilde{r}_A^2)\right]
 = \frac{\sin(\lambda)}{\lambda} \frac{4 \pi L_p^2}{3}(\rho+3p).
\end{eqnarray}
Multiplying both sides of the above equation by the same factor $2\dot{a}a$, after
using the continuity equation (\ref{Cont}), we get
\begin{eqnarray}\label{Frgb2}
\frac{d}{dt} \left\{ \left[\frac{\cos(\frac{\lambda}{G} \tilde{r}_A^2)}{\tilde{r}_A^2} + \frac{\lambda}{G} Si(\frac{\lambda}{G}\tilde{r}_A^2)\right] a^2 \right\}
= \frac{\sin(\lambda)}{\lambda} \frac{8 \pi L_p^2}{3} \frac{d}{dt} \left(\rho a^2 \right).
\end{eqnarray}
As in the previous section, we take the limit, $\sin(\lambda)/\lambda\simeq1$. Integrating Eq.(\ref{Frgb2}), yields
\begin{eqnarray}\label{Fried final 2}
\frac{\cos(\frac{\lambda}{G}\tilde {r}_{A}^2)}{\tilde {r}_{A}^2} +\frac{\lambda}{G} Si(\frac{\lambda}{G}\tilde {r}_{A}^2)
= \frac{8\pi L_p^2}{3}\rho,
\end{eqnarray}
where we have set the integration constant to zero.
This is the modified Friedmann equation for q-deformed entropy based on the law of emergence.
This equation (\ref{Fried final 2}) is found to be consistent with the one (\ref{Fried final}) derived from the unified first law of thermodynamics in Sect.\ref{FirstLaw}.
The result supports the viability of Padmanabhan's perspective on emergence gravity and its modified formula given by Sheykhi, and reveals that the law of emergence is associated with the unified first law of thermodynamics deeply at the apparent horizon of a FRW universe.
Incidentally, the approach we used is enough to reach the Friedmann equation of the FRW universe with any spatial curvature.

\section{The law of emergence and the horizon entropy maximization}\label{Max}
In this section, we are going to examine the maximization of q-deformed horizon entropy (\ref{q entropy}) in a $(3+1)$-dimensional FRW universe. Or rather, we investigate the constraints for q-deformed horizon entropy maximization.

\subsection{Horizon entropy maximization of q-deformed entropy}
It is widely recognized that every ordinary and isolated macroscopic system ultimately progresses toward an equilibrium state characterized by maximum entropy \cite{Max1},
\begin{eqnarray}\label{cond1}
 \dot S \geq 0, \, \textrm{always},
\end{eqnarray} and
\begin{eqnarray}\label{cond2}
   \, \, \, \, \, \,  \ddot S <0, \, \, \, \textrm{at least in the long run}.
\end{eqnarray}
In this formula, $S$ represents the total entropy of the universe which can be approximated as the horizon entropy, and the dots denote the derivatives with respect to the relevant variable, which is cosmic time.
In the context of Einstein gravity, our universe behaves as an ordinary macroscopic system that proceeds to a maximum entropy state with the constraints in Eq.(\ref{cond1}) and Eq.(\ref{cond2}). Subsequently, Krishna and Mathew extended the procedure to Gauss-Bonnet and Lovelock gravities for a spatially non-flat universe \cite{Max3,Max4,Max5}.
In this note, the process is extended to the q-deformed horizon entropy.

The q-deformed entropy proposed by Jalalzadeh for a quantum-deformed black hole is assumed to hold for the apparent horizon of the FRW universe, and is expressed as \cite{Quan0}
\begin{eqnarray}\label{qA entropy}
S_q = \pi\sin\left(\frac{A}{8G\mathcal N} \right)/\sin\left(\frac{\pi}{2\mathcal N} \right),
\end{eqnarray}
where $A = 4\pi \tilde{r}_A^2$ is the area of the apparent horizon.

In order to judge whether the horizon entropy is getting maximized over the course of cosmic time, the rate of change of the horizon entropy should be considered \cite{Quan0}.
We take the derivative of Eq.(\ref{qA entropy}) with respect to the cosmic time to get
\begin{eqnarray}\label{ds 1}
\dot S_q =\frac{2\pi\lambda}{G} \frac{\cos(\frac{\lambda}{G} \tilde{r}_A^2)}{\sin(\lambda)} \tilde{r}_A \dot{\tilde{r}}_A .
\end{eqnarray}
To satisfy the constraint $\dot S_q\geq 0$, $\dot{\tilde{r}}_A$ should always be greater than or equal to zero. In this case, the entropy will never decrease.
On the basis of Eq.(\ref{radius}) and Eq.(\ref{Fried final}), along with the continuity equation, $\dot\rho+3H(\rho+p)=0$, the first derivative of $\tilde{r}_A$ is obtained as
\begin{eqnarray}\label{dr 1}
\dot{ \tilde{r}}_A = 4\pi G \frac{\tilde{r}_A^3 H(1+\omega)\rho}{\cos(\frac{\lambda}{G} \tilde{r}_A^2)} ,
\end{eqnarray}
where $\omega$ is the parameter defined through the equation of state $p=\omega\rho$.
From recent observations, the universe is evolving to a pure de Sitter state with $\omega \geq -1$. Thus, it clearly shows $\dot{\tilde{r}}_A\geq 0$, which means that $\dot S_q$ is always non-negative and the constraint of $\dot S_q$ is met.

Next, our object is to check whether this entropy attains a maximum value in the long run characterized by the inequality $\ddot S_q< 0$.
Differentiating Eq.(\ref{ds 1}) with respect to the cosmic time, we get
\begin{eqnarray}\label{ds 2}
 \ddot S_q = \frac{2\pi}{G} \frac{\lambda}{\sin(\lambda)}
 \left\{ \dot{\tilde{r}}_A^2
 \left[\cos(\frac{\lambda}{G}\tilde{r}_A^2)-2\frac{\lambda}{G}\tilde{r}_A^2 \sin(\frac{\lambda}{G}\tilde{r}_A^2)\right]
 + {\tilde{r}_A}{\ddot{\tilde{r}}_A}\cos(\frac{\lambda}{G}\tilde{r}_A^2) \right\}.
\end{eqnarray}
Due to ${\dot{\tilde{r}}_A^2}$ being greater than or equal to zero, ${\ddot{\tilde{r}}_A}$ in the above equation should be less than zero so as to satisfy the constraint $\ddot S_q< 0$.
Therefore, the constraint for the horizon entropy maximization should be
\begin{eqnarray}\label{jrrj0}
  {\dot{\tilde{r}}_A^2}\left[\cos(\frac{\lambda}{G}\tilde{r}_A^2)-2\frac{\lambda}{G}\tilde{r}_A^2 \sin(\frac{\lambda}{G}\tilde{r}_A^2)\right]
  < - {\tilde{r}_A}{\ddot{\tilde{r}}_A}\cos(\frac{\lambda}{G}\tilde{r}_A^2),
\end{eqnarray}
in order to satisfy the constraint at least in the last stage. Within the limit of $0\leq{\frac{\lambda}{G}\tilde{r}_A^2}\leq\frac{\pi}{2}$, it can be found in Eq.(\ref{jrrj0}) that
\begin{eqnarray}
\cos(\frac{\lambda}{G}\tilde{r}_A^2)-2\frac{\lambda}{G}\tilde{r}_A^2 \sin(\frac{\lambda}{G}\tilde{r}_A^2)
\leq \cos(\frac{\lambda}{G}\tilde{r}_A^2).
\end{eqnarray}
Then, Eq.(\ref{jrrj0}) is simplified to
\begin{eqnarray}\label{jrrj1}
  | {\dot{ \tilde{r}}_A^2} |
  < | {\tilde{r}_A}{\ddot{\tilde{r}}_A} | .
\end{eqnarray}
This is the constraint for the q-deformed horizon entropy maximization in the simplest form.
To verify this constraint, we take the derivative of Eq.(\ref{dr 1}),
\begin{eqnarray}\label{dr 2}
\ddot{ \tilde{r}}_A =\frac{\tilde{r}_A^3\rho}{\cos(\frac{\lambda}{G}\tilde{r}_A^2)}
\left\{ [\frac{3}{2}+\frac{\lambda}{G}\frac{Si(\frac{\lambda}{G}\tilde{r}_A^2)
\tilde{r}_A^2}{\cos(\frac{\lambda}{G}\tilde{r}_A^2)} ](1+\omega)^2 H^2
+\frac{2\lambda}{G}\tan(\frac{\lambda}{G}\tilde{r}_A^2)\tilde{r}_A\dot{ \tilde{r}}_A(1+\omega)H
\notag\right.
\\ \phantom{=\;\;}
\left.
+(1+\omega)\dot H+\dot \omega H
\right\}.
\end{eqnarray}
In the final de Sitter epoch, $\omega \to -1$, so that all terms containing $(1+\omega)$ in Eq.(\ref{dr 2}) vanish. Because of the negativity of $\dot \omega$, we can find the condition $\ddot{ \tilde{r}}_A<0$. In the asymptotic limit, $t\to \infty$, as $\omega \to -1$, there will be $\dot{ \tilde{r}}_A \to 0$ from Eq.(\ref{dr 1}). Then the constraint (\ref{jrrj1}) for horizon entropy maximization is valid in the final stage.
Thus, in the context of q-deformed entropy, the horizon entropy of a non-flat universe will never grow unbounded, which ensures the maximization of entropy.

\subsection{Relationship between the law of emergence and horizon entropy maximization}
In this subsection, our aim is to check whether the law of emergence leads to the maximization of q-deformed horizon entropy.
Combining Eq.(\ref{dVt1}) and Eq.(\ref{ds 1}), we can relate the increase in the effective volume of the universe and the rate of change of the entropy as
\begin{eqnarray}
 \frac{d\widetilde{V}}{dt} = 2 L_p^2 \tilde {r}_A \dot S_q.
\end{eqnarray}
Then, Eq.(\ref{dV1}) for the law of emergence can be expressed as
\begin{eqnarray}\label{ds1NsNb}
 \dot S_q ={\frac{H}{2}(N_{\rm{sur}}- N_{\rm{bulk}})}.
\end{eqnarray}
From the relations in Eq.(\ref{Nsur2}), Eq.(\ref{Nbulk}) and Eq.(\ref{Frgb1}), the holographic discrepancy, $ (N_{\rm{sur}}-N_{\rm{bulk}}) $, is calculated as
\begin{eqnarray}\label{NsNb}
N_{\rm{sur}}-N_{\rm{bulk}} = \frac{4\pi \lambda}{G} \frac{\cos(\frac{\lambda}{G}\tilde {r}_A^2)}{\sin(\lambda)}\frac{\tilde {r}_A \dot{\tilde {r}}_A}{H}.
\end{eqnarray}
Since we assume the universe to be asymptotically de Sitter, $\dot{\tilde{r}}_A$ is greater than or equal to zero. As a result, the holographic discrepancy is always non-negative, which ensures the constraint $\dot S_q \geq 0$.

Next, to find out whether the horizon entropy is getting maximized, the second derivative of horizon entropy from Eq.(\ref{ds1NsNb}) is obtained as
\begin{eqnarray}\label{ds2NsNb}
 \ddot S_q ={\frac{\dot H}{2}(N_{\rm{sur}}- N_{\rm{bulk}})}+{{\frac{H}{2}} \frac{d}{dt}(N_{\rm{sur}}- N_{\rm{bulk}})}.
\end{eqnarray}
During the final de Sitter stage, the degree of freedom of bulk denoted as $N_{\rm{bulk}}$ tends towards $N_{\rm{sur}}$, thereby causing the first term, which involves ${(N_{\rm{sur}}- N_{\rm{bulk}})}$ in the aforementioned equation, to nullify.
The holographic discrepancy between $N_{\rm{bulk}}$ and $N_{\rm{sur}}$ will decrease closely to zero in the final stage, but it is always positive. That is,
\begin{eqnarray}\label{dNsNb0}
 \frac{d}{dt}(N_{\rm{sur}}- N_{\rm{bulk}})<0 ,
\end{eqnarray}
which ensures the negativity of $\ddot S_q$ in the long run.
Differentiating Eq.(\ref{NsNb}), we get
\begin{eqnarray}\label{dNsNb}
\frac{d}{dt}(N_{\rm{sur}}- N_{\rm{bulk}})
= \frac{4\pi}{G H} \frac{\lambda}{\sin(\lambda)}\left\{
\dot{\tilde{r}}_A^2 \left[\cos(\frac{\lambda}{G}\tilde{r}_A^2)
-2\frac{\lambda}{G}\tilde{r}_A^2 \sin(\frac{\lambda}{G}\tilde{r}_A^2)\right]
\notag\right.
\\ \phantom{=\;\;}
\left.
+ {\tilde{r}_A}{\ddot{\tilde{r}}_A}\cos(\frac{\lambda}{G}\tilde{r}_A^2)\right\}.
\end{eqnarray}
Combining Eq.(\ref{ds2NsNb}), Eq.(\ref{dNsNb0}) and Eq.(\ref{dNsNb}), we reach the constraint for the negativity of $\ddot S_q$,
\begin{eqnarray}\label{jrrj2}
  | {\dot{ \tilde{r}}_A^2} |
  < | {\tilde{r}_A}{\ddot{\tilde{r}}_A} | .
\end{eqnarray}
It is the same constraint as the one in Eq.(\ref{jrrj1}) derived from the maximization of q-deformed horizon entropy in the previous subsection.
Therefore, it can be concluded that the law of emergence leads to the maximization of q-deformed horizon entropy.

\section{Conclusions and discussions}\label{Con}
Based on the quantum deformation approach to quantum gravity, Jalalzadeh proposed a new entropy formula for black holes \cite{Quan1}. According to various quantum gravity proposals, this entropy is simplified in Ref.\cite{Quan0}, and is defined as the quantum-deformed entropy given by Eq.(\ref{q0 entropy}).
In this paper, based on the law of emergence of cosmic space proposed by Padmanabhan, the modified Friedmann equation from q-deformed entropy is derived in a $(3+1)$-dimensional case.
Work by Sheykhi \cite{Sheykhi5} shows that the unified first law of thermodynamics on the apparent horizon can be reformulated as a modified Friedmann equation of a FRW universe with any spatial curvature.
Assuming that the q-deformed entropy is associated with the apparent horizon of a FRW universe, we start with the unified first law of thermodynamics, ${dE = T_q dS_q + WdV}$, on the apparent horizon.
On the one hand, we derive the modified Friedmann equation from q-deformed entropy in a FRW universe by employing the unified first law of thermodynamics.
On the other hand, considering the proposal of the emergence of cosmic space, we obtain the modified Friedmann equation from q-deformed entropy by applying the modified law of emergence. The latter equation obtained coincides with the one derived from the unified first law of thermodynamics. It demonstrates the deep relevance between the unified first law of thermodynamics and the law of emergence at the apparent horizon of a FRW universe. These studies show that, in a sense, the unified first law of thermodynamics can be considered the origin of the emergence of cosmic space.

Pavon and Radicella have shown that our universe, with a history of Hubble expansion, emerges as an ordinary macroscopic system that evolves into an equilibrium state with maximum entropy \cite{Max2}.
After that, Krishna and Mathew examined the fact that the generalized holographic equipartition under Einstein, Gauss-Bonnet and Lovelock gravities is consistent with the constraints of entropy maximization.
In this research, we investigate the consistency of the law of emergence with the maximization of q-deformed horizon entropy. Our results show that the law of emergence and the horizon entropy maximization lead to the same constraints in the final. An ordinary and isolated macroscopic system will ultimately progress toward an equilibrium state characterized by maximum entropy, even in a cosmological context. In a sense, the emergence of cosmic space can be regarded as a tendency to maximize the horizon entropy in a non-flat universe.

All in all, the results of the modified Friedmann equation obtained in this note support the proposal of Padmanabhan \cite{Padmb3} and its modified version \cite{Sheykhi6}. In the context of q-deformed horizon entropy, we further reveal a profound connection between the law of emergence and the thermodynamics of horizon entropy.

We would like to further extend the research to higher-dimensional universes to discuss whether the law of emergence is consistent with the constraints of q-deformed horizon entropy maximization in the $(n+1)$-dimensional case. Perhaps this is a significant issue to investigate.

\begin{acknowledgements}
This work is supported by scientific research project of department of education of Hunan Province, No. 19C1895.
\end{acknowledgements}

\end{document}